\documentclass[]{spie}  %>>> use for US letter paper
\usepackage{amsmath,amsfonts,amssymb}
\usepackage{graphicx}
\usepackage[colorlinks=true, allcolors=blue]{hyperref}

\title{The Infrared Imaging Spectrograph (IRIS) for TMT: final software design update}

\author[a]{Edward L. Chapin$^*$}
\author[a]{Jennifer Dunn}
\author[b]{Takashi Nakamoto}
\author[c]{Jiman Simon Sohn}
\author[d]{Arun Surya}
\author[c]{Chris Johnson}
\author[d]{Shelley Wright}
\author[d]{Andrea Zonca}
\author[a]{David Andersen}
\author[e]{Eric Chisholm}
\author[e]{Kim Gillies}
\author[b]{Yutaka Hayano}
\author[a]{Glen Herriot}
\author[a]{Dan Kerley}
\author[c]{James Larkin}
\author[b]{Ryuji Suzuki}

\affil[a]{National Research Council Herzberg, 5071 W Saanich Rd, Victoria, V9E 2E7, Canada}
\affil[b]{National Astronomical Observatory of Japan, 2-21-1 Osawa, Mitaka, Tokyo, 181-8588, Japan}
\affil[c]{Department of Physics and Astronomy, Univ. of California, Los Angeles, CA 90095-1547, USA}
\affil[d]{Center for Astrophysics and Space Sciences, Univ. of California, San Diego, La Jolla, CA 92093, USA}
\affil[e]{Thirty Meter Telescope International Observatory, 100 W Walnut St, \#300, Pasadena, CA 91124, USA}

\authorinfo{$^*$ed.chapin@nrc-cnrc.gc.ca, phone +1 250 363 6971}

\begin{document}
\maketitle

% -----------------------------------------------------------------------------
% ABSTRACT
% -----------------------------------------------------------------------------

\begin{abstract}
  The InfraRed Imaging Spectrograph (IRIS) is the first-light client instrument for the Narrow Field Infrared Adaptive Optics System (NFIRAOS) on the Thirty Meter Telescope (TMT). Now approaching the end of its final design phase, we provide an overview of the instrument control software. The design is challenging since IRIS has interfaces with many systems at different stages of development (e.g., NFIRAOS, telescope control system, observatory sequencers), and will be built using the newly-developed TMT Common Software (CSW), which provides framework code (Java/Scala), and services (e.g., commands, telemetry). Lower-level software will be written in a combination of Java and C/C++ to communicate with hardware, such as motion controllers and infrared detectors. The overall architecture and philosophy of the IRIS software is presented, as well as a summary of the individual software components and their interactions with other systems.
\end{abstract}

% Include a list of keywords after the abstract
\keywords{instrument software, motion control, infrared detectors, instrument rotators, atmospheric dispersion correctors, adaptive optics, environmental control}

% -----------------------------------------------------------------------------
% INTRODUCTION
% -----------------------------------------------------------------------------

\section{INTRODUCTION}
\label{sec:intro}  % \label{} allows reference to this section

The InfraRed Imaging Spectrograph (IRIS) is a combined near-infrared (0.8 -- 2.4\,$\mu$m) imaging camera and Integral Field Spectrograph (IFS), with an AO-corrected $34'' \times 34''$ field of view provided by the Narrow Field Infrared Adaptive Optics System (NFIRAOS). As the first-light science instrument for the TMT, and one of the first TMT subsystems for which the final software design will be completed, IRIS will serve as a reference for future generations of TMT instruments. During the IRIS final design phase (2017-2021) the TMT Common Software (CSW) project was completed, which provides framework code and associated services (e.g., commands, telemetry) which instrument software components use. The TMT Executive Software (including observatory sequencers), as well as the telescope control system (TCS) and Adaptive Optics Executive Software (AOESW) have also advanced, requiring continual communication with their design teams in order to develop Interface Control Documents (ICD). Finally, the IRIS design itself has matured substantially since its Preliminary Design Review, with new technology choices for components such as the On-Instrument Wavefront Sensor detectors, and some of the motors and motion controllers. IRIS is being designed by an international consortium, including Herzberg Astronomy and Astrophysics (Canada); UCLA, UCSC, UCSD and Caltech (USA); NAOJ (Japan) and NIAOT (China). An overview of the complete instrument is given elsewhere\cite{larkin2020}.

\begin{figure}[ht]
  \begin{center}
  \includegraphics[width=\linewidth]{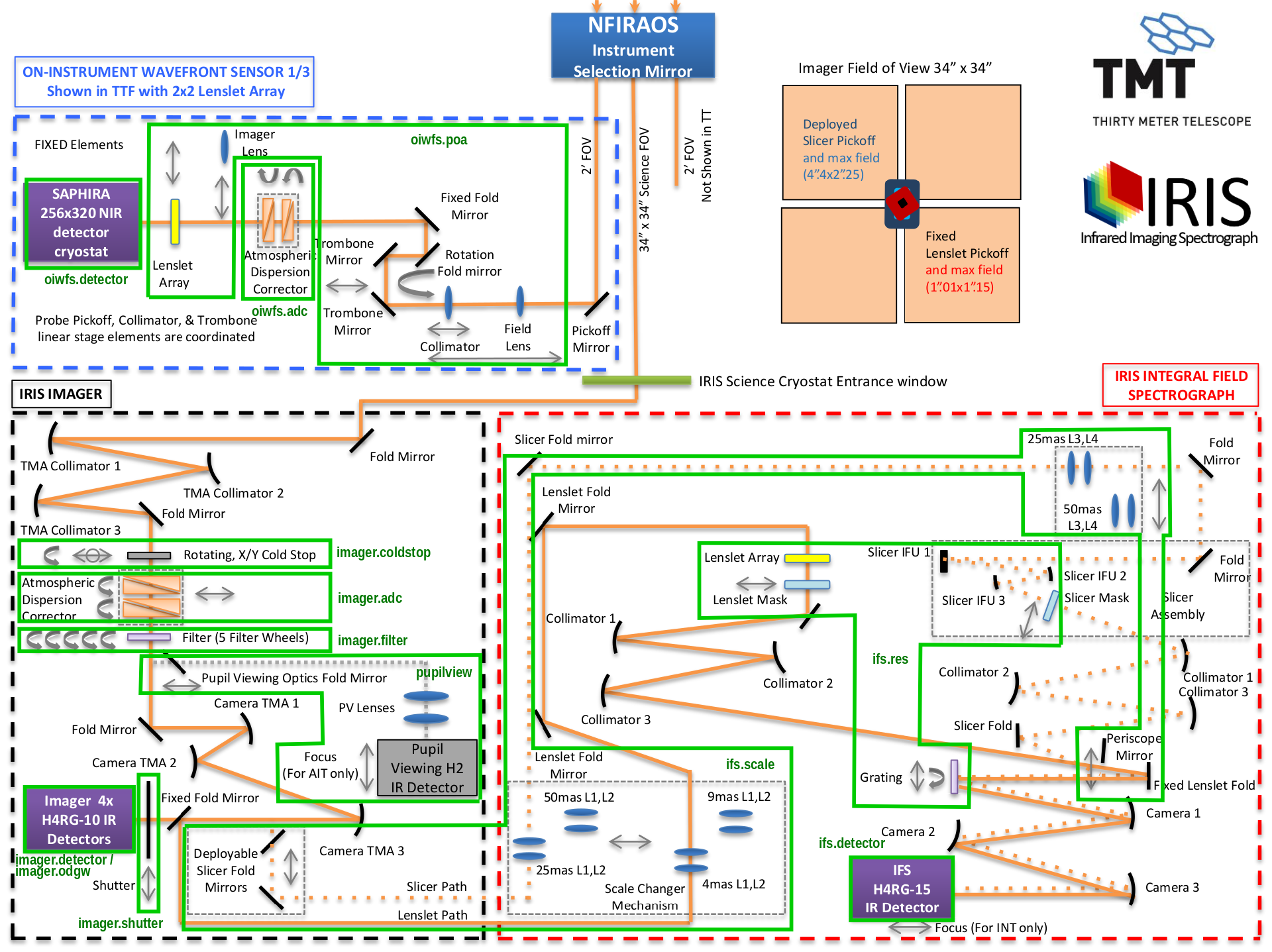}
  \end{center}
  \caption{\label{fig:blockDiagram}
  IRIS functional diagram. Light enters from NFIRAOS at the top of the figure, passes down through the On-Instrument Wavefront Sensor (OIWFS) chamber, which shares the NFIRAOS -30\,C environment, and then enters the science cryostat through a window at the bottom of the OIWFS where it illuminates an Imager, a Pupil Viewing Camera (to fine-tune the position of the Cold Stop) using movable pickoff optics, and an Integral Field Spectrograph (IFS) via a deployable fold mirror (with both image slicer and lenslet array modes, indicated by the dotted and solid brown lines, respectively). Moving components are indicated with grey arrows. All of these mechanisms and the infrared detectors are controlled by software ``assemblies'' as indicated by the green boxes, with names given in a dark green font. The following assemblies are not included in this diagram: Rotator, Science Cryostat Environmental Control (sc.cryoenv), Electronics Rack Power (el.power), and Electronics Rack Environment (el.env).
  }
\end{figure}

In this paper we summarize the overall architecture of the IRIS control software, and the primary interfaces to external systems. We also discuss our approach to using TMT CSW based on our prototyping experiences and participation in TMT Observatory Software reviews. Details of the IRIS Data Reduction System (DRS) are described elsewhere\cite{surya2020}. Finally, we give an overview of how individual components will be controlled, including:

\begin{itemize}
    \item four types of Infrared detectors: two different varieties of Teledyne $4096 \times 4096$ pixel HAWAII-4RGs (H4RG) for the science imager and IFS, and a $2048 \times 2048$ pixel HAWAII-2 (H2) for the Pupil Viewing Camera; and First Light Advanced Imagery C-RED One (CRED1) infrared cameras for the On-Instrument Wavefront Sensors (OIWFS);
    \item Atmospheric Dispersion Correctors (ADC) for both the OIWFS and Science Imager;
    \item a large direct-drive torque motor for the instrument rotator;
    \item numerous actuators that will operate both within the -30 C environment of the OIWFS, and cryo-cooled mechanisms within the science cryostat;
    \item environmental control of the OIWFS (ambient air pressure, but cooled to the -30 C temperature of NFIRAOS); and
    \item environmental control of the science cryostat.
  \end{itemize}

% -----------------------------------------------------------------------------
% ARCHITECTURE
% -----------------------------------------------------------------------------

\section{Software Architecture}
\label{sec:hierarchy}

The IRIS control software is built using TMT Common Software (CSW) framework code and services. The overall TMT observatory software architecture\cite{gillies2020} is split into a five-layer hierarchy consisting of: Physical hardware (Layer 0), Hardware Control Daemons (HCDs) -- the software used to provide low-level interfaces to hardware (Layer 1), Assemblies -- software that exposes an operationally-focussed higher-level interface to components (Layer 2), Sequencers -- used to coordinate the activities of collections of Assemblies (Layer 3), and User-facing applications (Layer 4). Communication between different software components (residing in any of the layers $\ge1$) is primarily accomplished using three different services: (i) the peer-to-peer Command Service; (ii) a single publisher / multiple subscriber Event Service for broadcasting telemetry via a central broker; and (iii) a peer-to-peer state publisher / subscriber service (PubSub service henceforth), mainly used by HCDs to report time-varying state information directly to the assemblies that control them. The information contained in commands, events, and PubSub states consist of messages with arbitrary numbers of key-value pairs. Additionally, a standard Configuration Service stores dynamic information to configure software components. Finally, an Alarm Service is provided for the purpose of bringing problems to the attention of a human operator. Each TMT software component registers a list of alarms that it maintains with the service, and periodically updates them. Some alarms must be continually refreshed within a certain period of time, serving as a ``heartbeat'' for the component; if such an alarm is not updated (even just to reaffirm its current state), it will appear in an alarm monitoring application.

Framework code\footnote{https://tmtsoftware.github.io/csw/} is provided to developers to write software for TMT and to communicate with each other via those services. The framework itself is written in Scala, a multi-paradigm (functional/object-oriented) language that targets the Java Virtual Machine (JVM), and is based around the Akka toolkit which implements actor-based concurrency. Briefly, actors are objects that communicate with each other using immutable messages (e.g., avoiding shared state variables), which are normally processed in the order that they arrive (though this can be configured to allow message priority). It is generally useful to reason about actors in terms of the types of messages that they receive, and their responses to them. Full access to the CSW services is available from Java and Scala, primarily as a series of Akka actor base classes (or traits in the case of Scala) that are extended by instrument developers. Partial CSW service client support is also offered in C/C++ and Python at the time of writing. The IRIS team is developing most of the control software in Java, with some low-level portions of HCDs written in C. The Data Reduction System (DRS) (discussed in a separate paper) is being implemented in Python 3.

The IRIS software team has gained experience using TMT CSW through collaboration with the observatory on prototyping efforts throughout the CSW build phase which has now concluded. IRIS software team members have also participated in TMT Observatory Software reviews, and \emph{vice versa}. We believe there is now a good mutual understanding about how IRIS instrument software will fit into the overall vision for the observatory, which will be key to its success.

% 
% Components
%

\subsection{Decomposition into components}

\begin{figure}[ht]
  \begin{center}
  \includegraphics[width=\linewidth]{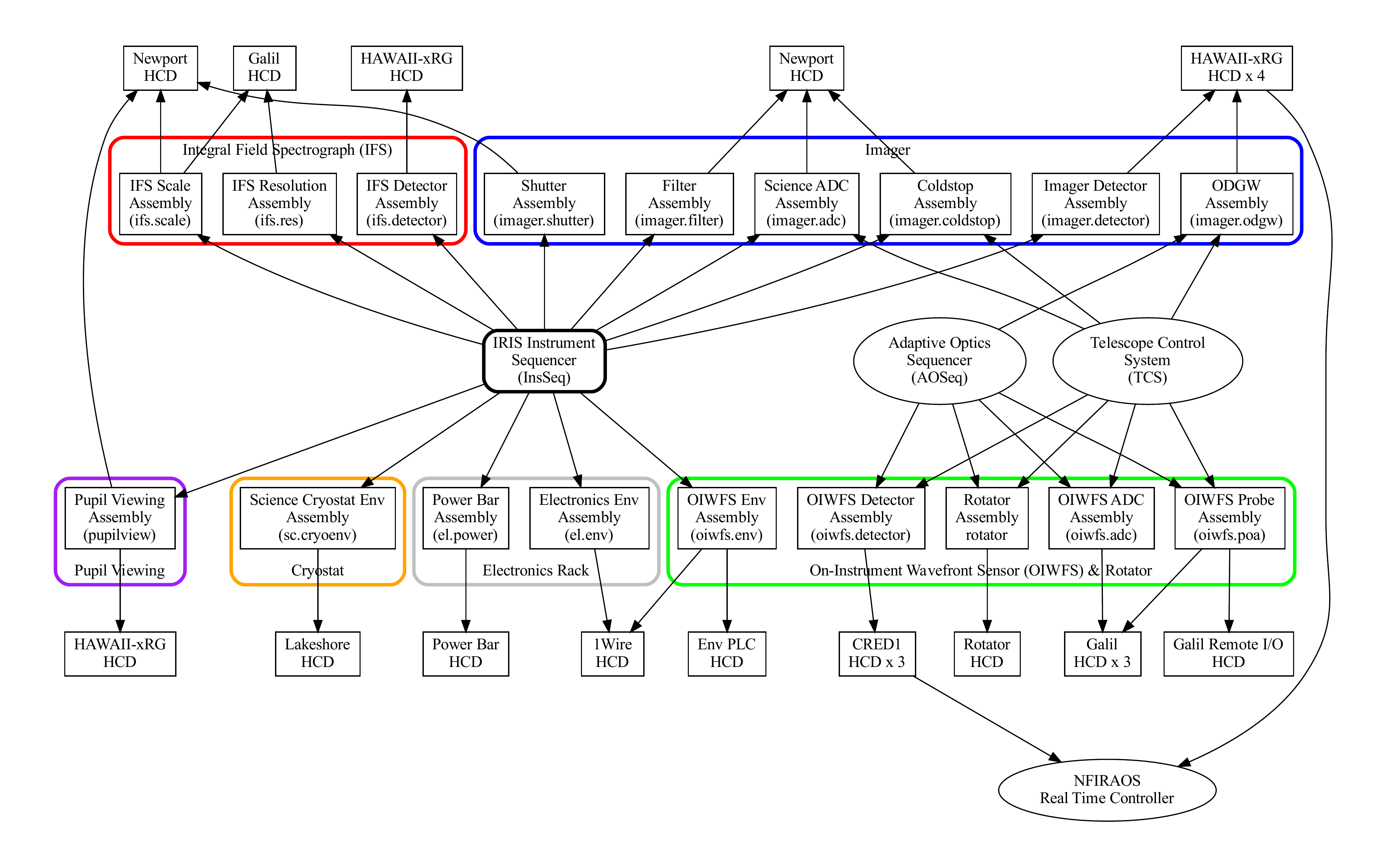}
  \end{center}
  \caption{\label{fig:context}
  IRIS software context diagram. Sequencers and the TCS are placed at the middle, with the assemblies (grouped into related clusters) around them. Finally, HCDs and the external interface for wavefront sensor pixels sent to the NFIRAOS RTC lie on the outside of the diagram. The assemblies can be located by name in Figure~\ref{fig:blockDiagram}.
  }
\end{figure}

The instrument software discussed in this paper can be decomposed into individual components that belong to Layers 1--3 of the TMT hierarchy. Low-level HCDs are designed to interface will all of the mechanisms (motion controllers, detectors, etc.) that have been selected by the IRIS team. The HCDs, in turn, are controlled by a number of assemblies that correspond to recognizable high-level components of the instrument (e.g., instrument rotator assembly, science ADC assembly, IFS detector assembly) and provide command interfaces that are meaningful to observatory operations (e.g., move OIWFS probes to particular x,y coordinates, take exposure with detector). 
Finally, an IRIS instrument sequencer (InsSeq) is used to coordinate the activities of the instrument. Another interesting feature of the IRIS software design is the fact that some assemblies are under the control of the observatory Adaptive Optics (AO) system (instrument rotator, On-Instrument Wavefront Sensors, On-Detector Guide Windows), and are configured/commanded via the AO Sequencer (AOSeq) being developed by the TMT AO team, rather than the InsSeq. The coordination of the instrument with the AO system and the Telescope Control System (TCS) are the subject of an earlier paper\cite{chapin2018}.

Figure~\ref{fig:context} shows the IRIS software components: the Instrument Sequencer is placed at the middle; assemblies lie around it grouped into physically related clusters (large colored boxes); and HCDs (rectangles) are at the periphery. The dot prefix names for the assemblies are included in brackets for reference. The entities in ovals represent the main external software connections: the AO Sequencer configures and sequences the activities of all observatory AO components, the Telescope Control System provides time-varying position demands for numerous mechanisms in the instrument as the telescope position changes, and the NFIRAOS Real Time Controller (NRTC) receives pixel streams directly from the OIWFS and On-Detector Guide Window (ODGW) detector HCDs over dedicated high-speed / low-latency Ethernet connections.

% 
% Common Design Patterns
%

\subsection{Common Design Patterns}

The IRIS team, in collaboration with the NFIRAOS team (due to substantial overlap of personnel at HAA) has created a set of design principles to help simplify the development of individual software components. Central to this Common Design Patterns (CDP) document is the idea that any portion of the software that will function independently will be represented by a state machine, referred to as a ``functional group''. The state itself is expressed as a multi-dimensional tuple, with the particular set of axes depending on the nature of the group. The following points summarize the main features of this model:

\begin{itemize}
  \item a functional group can only execute one command at a time (though long-running commands can be stopped or pre-empted),
  \item each functional group has an associated ``state tuple'',
  \item the state tuple is published as an event each time it changes, for the benefit of external systems,
  \item the attributes (axes) of the state tuple are built up from a standard set representing a range of typical behaviors (e.g., one representing command execution status, another for stages that can move continuously etc.), and
  \item many command names are standardized to match the associated behavior of the state model.
\end{itemize}

Assemblies and HCDs will have one or more functional groups, depending on their complexity. For example, the Science ADC (Section~\ref{sec:imagerAdc}) has one functional group representing control of its retraction mechanism, while another functional group controls its wedge prism rotation angles. This decomposition allows it to simultaneously act on commands to deploy the ADC and to rotate its prisms into position. In contrast, the Rotator assembly will only have a single functional group, and can generally only execute a single command at a time.

In addition to the underlying functional group model for software components, various other areas are covered by the CDP. For example, the conversion between device-specific units and physical units (e.g., from motor counts to millimeters) is the responsibility of HCDs. However, this conversion is generally at most a linear (gain and offset) relationship. If a more complex pointing model is required it is implemented in assemblies. Another clear delineation between the roles of HCDs and Assemblies is the implementation of simulations. Whenever possible, hardware simulations will be implemented at a low level within the HCDs. While this approach may be more labour intensive than simply ``throwing a switch'' in an assembly to change to a code branch that replaces the HCD with simulated I/O, it ensures that a much larger portion of the operational software will be executed in simulated scenarios.

Finally, it is expected that many features of the CDP will be codified as a library during the build phase. For example, Java base classes that represent the functional groups, and the commands and state tuple events that go with them will assist with adherence to the model, and reduce the overall development time. It is hoped that this strategy will improve code readability and maintainability once it is delivered to the observatory.

\begin{figure}[ht]
  \begin{center}
  \includegraphics[width=0.7\linewidth]{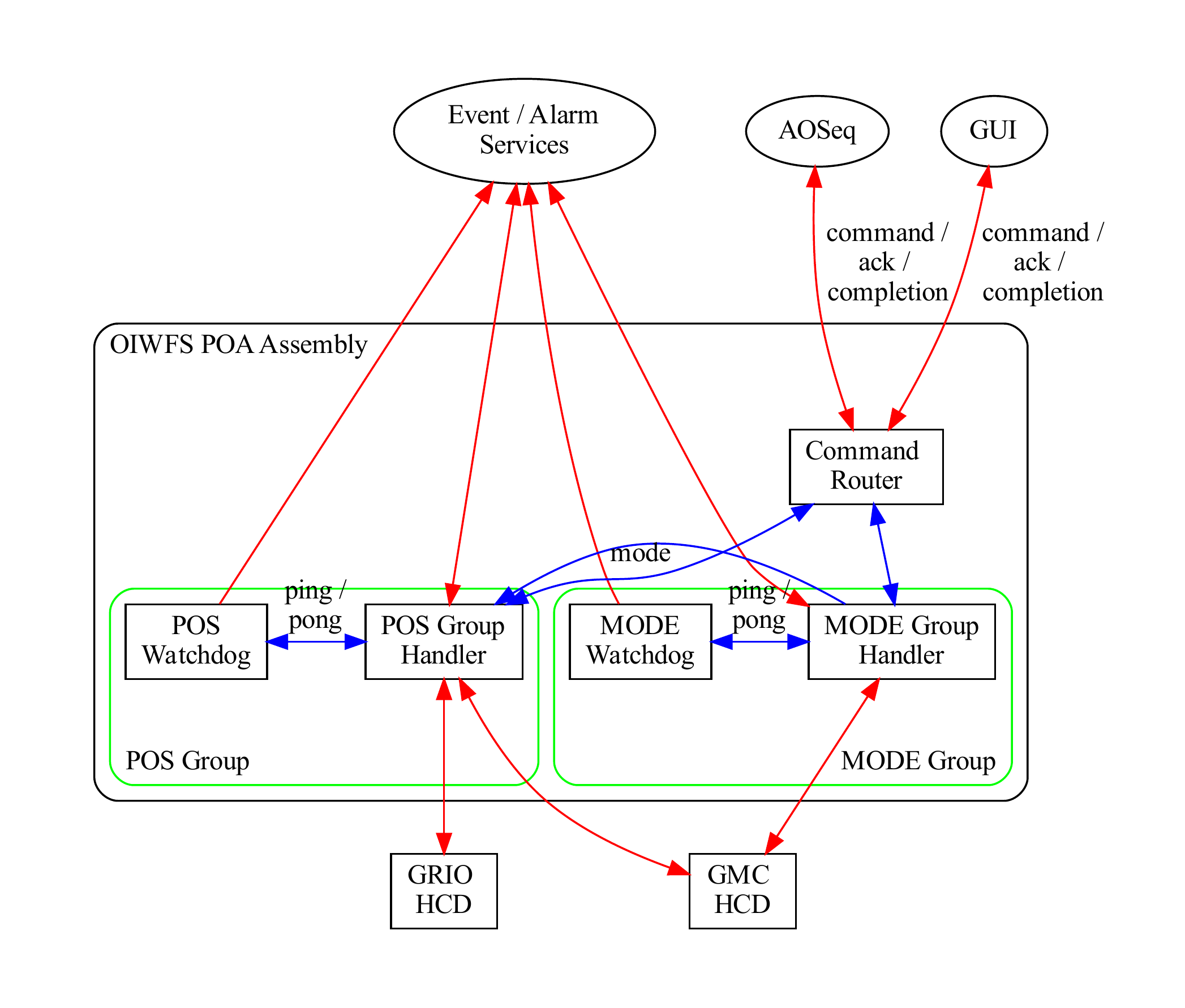}
  \end{center}
  \caption{\label{fig:poa_design}
  Simplified actor design for the OIWFS Probe Arm Assembly (Section~\ref{sec:oiwfsPoa}), consisting of two functional groups, using Common Design Patterns.
  }
\end{figure}

Figure~\ref{fig:poa_design} shows a simplified actor design for the OIWFS Probe Arm Assembly (discussed further in Section~\ref{sec:oiwfsPoa}) which involves two functional groups, one for positioning the arms in the OIWFS focal plane, and the other for swapping optics to change the mode of operation (the real design will have additional actors, including CSW interfaces). A ``command router'' is the point of entry for commands coming from the AOSeq, and it forwards them to the ``Group Handler'' for the respective functional groups. The Group Handlers maintain the group state machines, and at all times are able to determine the validity of a command. The ``Watchdog'' actors periodically send \emph{ping} messages to the Group Handlers, to which they expect a \emph{pong} response on a short time scale (e.g., less than a few seconds). If it does not, it raises an alarm to indicate that the group is not responsive.

The following lists a subset of the messages acted upon by the POS (positioning) group handler to illustrate how things work:

\begin{itemize}
  \item \textbf{follow}: start following positions from the TCS (command from the AOSeq via the Command Handler).
  \item \textbf{oiwfsProbeDemands}: update target positions for the probes (event published by the TCS and relayed by the Event Service), received at 20\,Hz.
  \item \textbf{gmcState}: Galil motion controller state (position) updates, provided at $\sim$100\,Hz (using state PubSub system). Each time this is received, the Group Handler responds by deriving new demands for the controller at the current instant, based on the most recent positions provided by the TCS and a path-following algorithm to avoid collisions, and sends them to the GMC HCD using a command. If a certain amount of time has elapsed, a new state event for the assembly including the new demands will be published.
  \item \textbf{grioState}: The Galil Remote I/O state (switches) updates indicate whether collisions have occured. If they have, the Group Handler updates its published state event to indicate that a collisions has occurred, and also raised an alarm. At this point an operator would be informed of the problem, and they can issue commands via a GUI to recover from the collision. If a certain amount of time has elapsed, a new state event for the assembly including the new demands will be published.
  \item \textbf{ping}: the Group Handler receives these messages from the Watchdog (e.g., at $\sim$1\,Hz), and must respond with a \emph{pong} message to indicate that it is responsive.
\end{itemize}

The various actors, commands, and state messages will be derived from base classes in the CDP library.

\section{Instrument Sequencer}
\label{sec:sequencer}

The IRIS Instrument Sequencer accepts high-level commands from the observatory (such as the desired instrument configuration for the next observation, usually from the top-level Observatory Control System sequencer), and translates them into lower-level commands for individual IRIS assemblies (excluding those under control of the AO Sequencer: oiwfs.adc, oiwfs.poa, rotator). The sequences themselves, with support for both linear and parallel operations, are described in a Domain Specific Language (DSL) which is currently being implemented in Kotlin by TMT. The intent is for non-software engineers to be able to modify them as needed to support observatory operations (e.g., no re-compilation nor complex software deployment should be needed). At the time of writing, most of the effort on the Instrument Sequencer has gone into the development of ICDs and a sequences document that lists all of the required tasks and operations that IRIS must be able to complete. It also demonstrates how specific commands for the assemblies will accomplish those tasks. TMT is currently building the sequencer framework\cite{weiss2020} that will be used to host the IRIS Instrument Sequencer.

% -----------------------------------------------------------------------------
% MOTION CONTROL
% -----------------------------------------------------------------------------

\section{Motion Control}
\label{sec:motion}

% 
% Ambient Mechanisms
%

\subsection{Ambient and Sub-zero Mechanisms}
\label{sec:ambient}

This section describes control software for mechanisms that operate both at ambient (dome) temperatures and pressures, and also at ``sub-zero'' temperatures. These latter components are cooled to the -30\,C temperature of the shared IRIS OIWFS and NFIRAOS environment.

Brushed DC servomotors will be used throughout the OIWFS, such as the probe arms (both the rotating and linear stages, as well as trombone and collimator stages), and the rotating ADC prism stages. Ethernet-accessible Galil DMC-40x0 controllers\footnote{https://www.galil.com/motion-controllers/multi-axis/dmc-40x0} will drive the motors, and can make use of both shaft and load encoders to minimize the impact of backlash.

A common Galil Motion Control (GMC) HCD is being designed to support the needs of IRIS and NFIRAOS. Each Galil controller can handle up to eight axes (though groups of four channels share a common amplifier module). There will therefore be multiple controllers with corresponding instances of the GMC HCD. Meeting some of the precise timing requirements with these controllers will be a challenge as there is no way to perform absolute synchronization between the controller clock and the servers that are sending the demands. Additionally, multiple assemblies will need to connect to a single GMC HCD instance (to send demands to particular axes under their control). Early prototyping\cite{chapin2018} showed that the controller can be configured to free-run at an internal rate of 100\,Hz, and for an internal control loop of the HCD to generate new demands for the controller each time it receives an update (i.e., it can run on a clock that is triggered by the 100\,Hz state updates). In parallel the HCD will receive demands asynchronously from assemblies. Interpolation and extrapolation of these demands evaluated at the instants they are to be sent to the controller by the inner loop should provide the required accuracy for the mechanisms with the tightest positioning requirements.

Precise calibration of the instrument rotator and OIWFS mechanisms described in this section will be derived from measurements of calibration sources within NFIRAOS.

% ROTATOR

\subsubsection{The IRIS instrument rotator (rotator)}
\label{sec:rotator}

The IRIS instrument rotator\cite{crane2020} compensates for sky rotation (due both to the varying parallactic angle and telescope elevation, since the instrument is on a Nasmyth platform). While the rotation rate will typically be modest, large speeds will be reached when transiting at high elevations (e.g., nearly 14 deg\,min$^{-1}$ when transiting 1\,deg from zenith). Furthermore, the accuracy of the rotation must meet or exceed 0.2\,deg RMS (though this requirement will be relaxed at lower zenith angles due to the high rates mentioned).

Unlike the other ambient mechanisms in IRIS which use brushed DC servomotors, the instrument rotator is based on a 1.2-m ETEL brushless DC direct-drive torque motor\footnote{https://www.etel.ch/torque-motors/overview/}, with a custom 1.8\,m diameter optical load encoder for feedback. A direct-drive solution, rather than a more standard geared approach was chosen to improve the angular accuracy of the system.

The rotator is (by far) the largest motor in IRIS. While it could conceivably be driven by a Galil Motion Controller like the other ambient mechanisms, the team is considering an ETEL controller recommended for the particular motor that we have selected. Despite the overhead of developing a new HCD, the ETEL controllers have additional features such as cogging control that may result in significant performance improvements. Time spent on debugging may also be reduced by using a motor and controller pair from the same manufacturer.

In addition to the rotator motor itself, the rotator assembly will control a second, smaller motor below IRIS for the Instrument Wrap. While the final design of the Instrument Wrap is pending, it will likely be driven by the same controller that is selected for the main rotator.

% OIWFS POA

\subsubsection{OIWFS Probe Positioners (oiwfs.poa)}
\label{sec:oiwfsPoa}

The OIWFS Probe Arm assembly controls the three movable probe arms (POA) for the OIWFS. Each arm is positioned in the focal plane with a rotational and linear extension stage, each with a shaft and load encoder. The arms also have fold mirrors (trombones) to maintain path length, and moving collimators to account for the changing distance of the OIWFS from the curved focal surface; both of these motors also have shaft and load encoders. Finally, there will be rotary deployment stages to switch between a single lens (TT), and 2x2 Shack Hartman (TTF) optics. These deployment stages have their end positions determined by hard stops, and the connection between the motor and the stage is facilitated by a spring; once initial contact between the stage and stop has been made the motor will be driven slightly past this position so as to load the spring and ensure that the optic does not move. As deployment does not require precise positioning, only shaft encoders will be used for feedback. Combined with the two axes required for the two ADC prisms (discussed in the next section), a single 8-axis Galil controller is used to control each of the three OIWFS probes.

As part of the AO system, the assembly is configured by the AOSeq. Position demands for the probes are provided by the TCS in the native IRIS coordinate system. As discussed in Section~\ref{sec:oiwfsAdc}, image offsets incurred by the ADCs are fed back to the TCS, and removed from the demands prior to their being sent. The trombone and collimator positions are internally determined from lookup tables based on the locations of the probes in the patrol area.

\begin{figure}[ht]
  \begin{center}
  \includegraphics[width=0.45\linewidth]{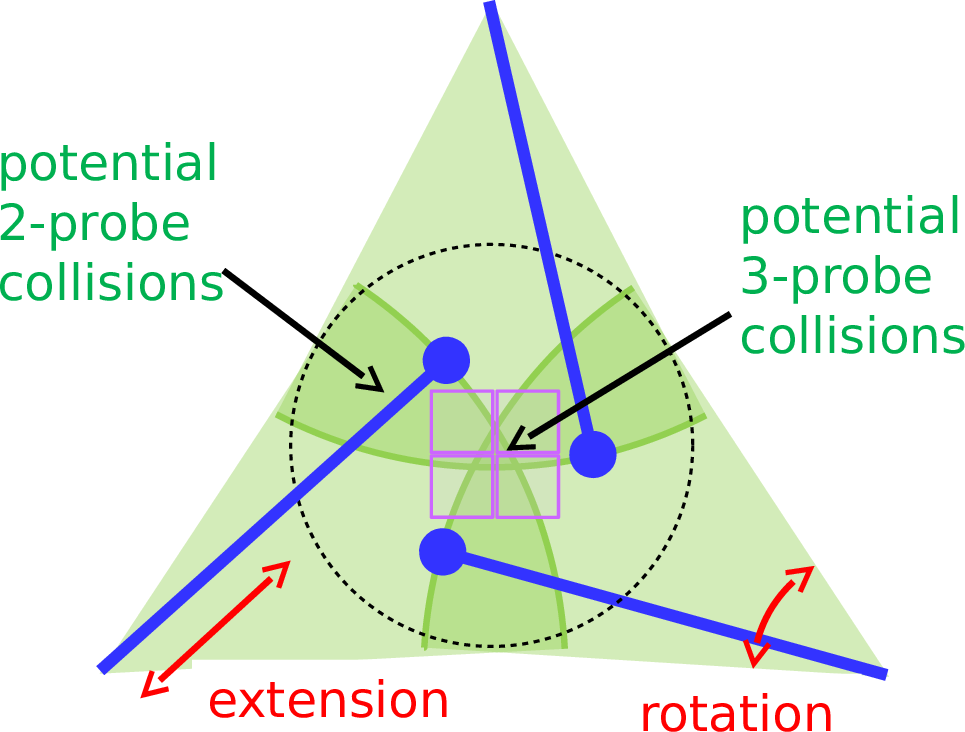}
  \end{center}
  \caption{\label{fig:probe_patrol}
  Illustration of the OIWFS probe patrol areas. Blue lines with circular heads represent the probes (rotation and extension stages), and the purple squares depict the footprint of the four IRIS imager chips. Individual probes can reach spaces accessible by adjacent probes in order to increase sky coverage. A small region at the center of the field is accessible by all three of them.
  }
\end{figure}

Probe positioning is a complex problem because the OIWFS has been designed to allow the arms to share much of the patrol region (to maximize sky coverage), including a small area at the centre of the field accessible by all three. The resulting possibility for collisions is illustrated in Figure~\ref{fig:probe_patrol}. The approach taken to collision avoidance in software is based on the calculation of ``virtual potential functions'', whereby the work space for each probe is represented with a height map. Peaks in this map indicate the locations of obstacles (other probes, the projected imager footprint), while the minimum is located at the target (guide star) position. Each step in the motion integration uses the local downhill gradient of this potential function to determine the direction of travel.

For normal observations (sidereal tracking), the guide star locations are quasi-static. The probes drive to their targets following the gradient descent method, and then remain stationary except for small adjustments. Any unused probes are parked at the edge of the patrol area. IRIS will also support non-sidereal tracking for observations of solar system bodies. For these cases guide stars will move with respect to the science target, possibly drifting out of the field (or close to the science target). In these cases individual probes may be reconfigured during an observation to (i) follow a new guide star that has entered the field of view; or (ii) park at the edge of the field until a new star becomes available. The individual probe reconfiguration problem is addressed by deactivating the repulsive potential for probes that are still tracking (i.e., so they remain fixed on their targets), and only activating the repulsive potential component for probes that are being steered to new targets (i.e., so that they move around the actively tracking probes). A more comprehensive description of the collision avoidance algorithm is the subject of a previous study\cite{chapin2016}. Timing and tracking requirements for IRIS mechanisms that following TCS demands have also been examined in an earlier paper\cite{chapin2018}.

In addition to the collision avoidance algorithm, the base of each probe is mounted using springs, so that in the event of a collision (e.g., due to motors losing their datum for some reason) they will flex to avoid damaging them. Switches on the base will sense this flexure (including the direction), and report this information back to the assembly from a Galil Remote I/O (GRIO) unit (with corresponding GRIO HCD), which will then halt motion and raise an alarm for the operator. Since the sign of the collision is known, the power to the rotational stage will only be deactivated in the direction that caused the collision, allowing the probes to ``back out'' as part of the recovery procedure.

Rather than moving the probes, the effective location of each OIWFS can also be varied by moving the readout windows on the respective detectors (Section~\ref{sec:oiwfsdetector}), so-called ``On-Chip Guiding''. When performing probe guiding the positioners will respond to changing TCS targets. Conversely, for on-chip guiding the readout windows move. The particular tracking mode is requested via a command to the oiwfs.poa assembly from the AOSeq, and coordination with the oiwfs.detector assembly is accomplished by sharing the current tracking mode and positions of the probes using the Event Service.

The trombone and collimator positions are calculated using lookup tables as a function of individual probe locations. The tables are based on modelling and calibration work performed as part of the optical design and testing\cite{atwood2020}.

A subset of OIWFS POA design is shown as an example in Figure~\ref{fig:poa_design}.

% OIWFS ADC

\subsubsection{OIWFS ADC Control (oiwfs.adc)}
\label{sec:oiwfsAdc}

This assembly controls the three OIWFS ADCs, each of which consists of two wedge-shaped prisms. Their differential rotation is used to vary the dispersion power of the ADC, while the combined (mean) rotation of the prisms tracks the projected elevation axis in the OIWFS focal plane. The rotation rates that the ADC must achieve are similar to the requirements for the rotator, since the horizon rotates at a comparable rate to that of the sky in the most extreme cases (transiting near zenith).

The goal of the ADC control problem is to minimize the smearing of astronomical targets, whose light is emitted at a range of wavelengths (their Spectral Energy Distribution, or SED), and which is dispersed along the elevation axis by the atmosphere. ADC control can be expressed as an optimization problem, in which the difference between a theoretical dispersion curve for the ADC and the atmospheric dispersion at the same instant (inferred from a model that takes into account the current telescope observing geometry and environmental measurements) is minimized. The algorithm is further enhanced by weighting the comparison by wavelength, taking into account the SED of the target, the filter bandpass, and detector quantum efficiency (QE). ADCs also induce a mean shift in the position of the light, an effect that must be accounted for.

\begin{figure}[ht]
  \begin{center}
  \includegraphics[width=0.45\linewidth]{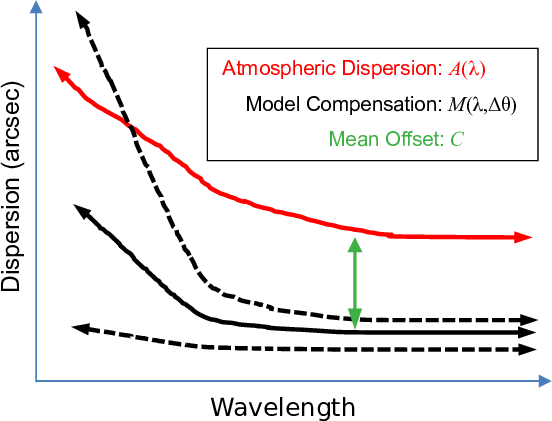}
  \end{center}
  \caption{\label{fig:adc_fitting}
  Illustration of ADC power setting optimization. The TCS provides a calculated value for the atmospheric dispersion $A(\lambda)$ (red). The oiwfs.poa assembly then chooses the power setting, based on differential prism rotation $\Delta\theta$ (black lines), that minimizes the mean square difference between its dispersion curve and that of the atmosphere (solid black line), including a wavelength weighting function (based on the source SED, filter bandpass, and detector QE), but ignoring the mean offset $C$ (green), as this shift is compensated for with a probe pointing offset.
  }
\end{figure}

The TCS will continuously publish dispersion events including the following attributes: (i) the orientation of the axis of dispersion, $\bar{\theta}$ (to account for sky rotation); (ii) a wavelength array, $\lambda_i$, at which the subsequent functions are sampled; (iii) the wavelength weight array to be applied to the dispersion curve, $W(\lambda_i)$, sampled at $\lambda_i$; and (iv) the current modeled atmospheric dispersion curve, $A(\lambda_i)$, sampled at $\lambda_i$. Note that if the TCS atmospheric dispersion model is updated by the observatory, it can simply evaluate the new model without requiring any changes to software interfaces (i.e., this is a more future-proof interface than having the TCS publish dispersion model parameters directly). Internally, the ADC assembly has an optical model for the ADC dispersion developed by the instrument team\cite{atwood2020}, and updated by on-sky calibration, which is represented as a 2-dimensional lookup table providing dispersion as a function of wavelength and differential prism rotation (power), $ M(\lambda,\Delta\theta)$. Since the ADC cannot perfectly counteract this atmospheric dispersion at all wavelengths, the solution is weighted toward the wavelengths of greatest interest by $W(\lambda)$.

The ADC power set so that it minimizes the mean square difference between the atmospheric dispersion, $A$, and ADC dispersion, $M$, with the supplied weights $W$. The following fitting function (involving sums over discrete wavelengths $\lambda_i$), which ignores the mean offset, is used:

\begin{equation}
E^2(\Delta\theta) = \sum_i { W(\lambda_i) \left[ A(\lambda_i) - M(\lambda_i,\Delta\theta) - C(\Delta\theta) \right] }^2
\end{equation}

Here $C(\Delta\theta)$ is the weighted mean offset (across wavelength) at the differential power setting $\Delta\theta$. It is calculated in the following way:

\begin{equation}
C(\Delta\theta) = \frac{ \sum_i{W^2(\lambda_i)A(\lambda_i)} - \sum_i{W^2(\lambda_i)M(\lambda_i,\Delta\theta)} } { \sum_i {W^2(\lambda_i)} }
\end{equation}

Given the dispersion, the OIWFS ADC assembly simply evaluates $E^2$ at a range of power settings $\Delta\theta$, choosing the power that minimizes its value, $\Delta\theta_\mathrm{min}$.

The orientation of the compensation is selected by rotating both ADC prisms such that their mean angle is parallel to the projected elevation axis, but applies a dispersion of opposite sign to that of the atmosphere. 

In the case of the OIWFS ADCs, the image shifts that they cause will appear as tip/tilt (TT) offsets to the adaptive optics system, which it will attempt to correct if they are not otherwise accounted for (thus shifting light sent downstream to the science dewar, and appearing as a pointing error in science exposures). To avoid this undesirable behavior, the oiwfs.poa assembly will publish the expected shifts, $C$, at all times for the benefit of the TCS. The TCS will, in turn, subtract these offsets in position demands subsequently sent to the OIWFS probes, so that NGS spots will appear in the OIWFS detectors where they are expected.

% 
% Cryogenic Mechanisms
%

\subsection{Cryogenic Mechanisms}

The Science Cryostat contains a number of mechanical actuators that are used for positioning optical components. These devices must operate at cryogenic temperatures (77\,K), and minimize heat production (which is difficult to dissipate in a vacuum). Cryogenic stepper motors are used for this purpose throughout the IRIS design. For mechanisms that require continuous motion (e.g., the Rotating Coldstop), prototyping efforts by NAOJ detected unacceptable heat generation when operated with Galil motion controllers. Newport XPS-D 8-axis motion controllers\footnote{https://www.newport.com/f/xps-d-universal-motion-controller} were later found to meet the requirements. A Newport controller HCD has not yet been designed, but initial investigations into its capabilities suggest that it will be possible to write one that functions in a similar manner to the Galil GMC HCD. Some of the cryogenic mechanisms that simply move a component into a position and stop will continue to use Galil controllers, since it is believed these stages can be de-energized upon completion of a move.

Position calibration for the steppers will be accomplished using a combination of mechanical lever limit switches, and Asahi Kasei Microdevices HG-106A hall sensors\footnote{https://www.akm.com/global/en/products/magnetic-sensor/hall-element/ga-as-low-drift/hg106a/}.

To protect against overheating, motion will be prevented if a count of recent steps exceeds some threshold. Additionally, all of the cold mechanism assemblies will receive temperature values published by the Cryostat Environment Assembly. If a stage temperature exceeds a configured threshold, the assembly immediately halts any active motion and powers off the motor to prevent damage.

% IMAGER COLDSTOP

\subsubsection{Rotating Cold Stop (imager.coldstop)}
\label{sec:coldstop}

The cold stop is the first moving component within the cryostat after the window. Located at a pupil position within the cryostat, it masks everything outside of an image of the primary mirror, and stops down the field to reduce noise created by the thermal emission of the surrounding telescope structure. Its shape matches the pupil, e.g. a serrated primary with a central obscuration and spider. As the instrument rotator stabilizes the image of the sky (Section~\ref{sec:rotator}), the projected pupil will also rotate, meaning that the cold stop will need to meet similar requirements to those of the rotator. Additionally, any ``wobble'' caused by the rotator and/or cold stop rotation stage will need to be compensated to ensure that pupil features are effectively masked at all times. This can be achieved by over-sizing the features in the mask, though relying completely on this approach is undesirable because it would substantially reduce the throughput of the instrument. For this reason the cold stop will be able to track continuously in X,Y coordinates, in addition to more modest over-sizing of the mask.

The cold stop therefore consists of one rotary stage, and X,Y linear stages for positioning, both driven by a Newport Controller. The precise position of the cold stop is calibrated from observations with the Pupil Viewing Camera (Section~\ref{sec:pupilview}). The X stage motion is further decomposed into an X-coarse stage that completely removes the cold stop from the field of view, which is used to enable taking of images with the Pupil View Camera (as well as images with the Imager and IFS without the stop, if desired). An additional X-precise stage (with a smaller range of motion) is used for fine adjustments. Two functional groups are planned for this assembly: one to control the X-precise stage, Y stage, and rotary stage (which operate in unison), and another to control the X-coarse deployment stage.

% IMAGER ADC

\subsubsection{Imager ADC (imager.adc)}
\label{sec:imagerAdc}

Following the cold stop is the Science ADC. It will operate under the same principle as the OIWFS ADCs (Section~\ref{sec:oiwfsAdc}), with rotary stages controlling a pair of wedge prisms. Like the cold stop, there will also be a separate linear deployment stage for moving the ADC in and out of the light path. All of these stages will be driven by Newport controllers. As such, the Imager ADC will consist of two functional groups (one for deployment, and the other for dispersion correction). The deployment stage for the ADC will be very similar to that of the Cold Stop, and likely use shared code.

Operationally, the main difference between the science ADC and the OIWFS ADC is that any shifts incurred by the ADC will move science targets about the image plane (i.e., causing a pointing error). In the case of the OIWFS, feedback of this expected shift from the assembly to the TCS is used to generated corrected position demands for the individual OIWFS probes. In contrast, the TCS will interpret shifts published by the imager.adc assembly as pointing corrections for the entire telescope.

Another distinction is that while the weights supplied to the ADC optimization algorithm for the OIWFS include the SEDs of guide stars (multiplied by the filter bandpass and detector QE), the TCS will provide the SED of the science target (if known, and again, multiplied by the filter and QE) to the imager.adc assembly .

% IMAGER FILTER

\subsubsection{Filter Selector (imager.filter)}
\label{sec:filter}

Five filter wheels are situated below the ADC. This assembly controls all of these wheels as a single unit (since, for example, four of the wheels must be set to a passthrough when a filter is selected on the remaining wheel), again using steppers driven by a Newport Controller. The assembly therefore requires only a single functional group.

% PUPILVIEW

\subsubsection{Pupil Viewing (pupilview)}
\label{sec:pupilview}

Below the filter wheel is a space where a pickoff mirror can be deployed using a linear stage to direct light to the pupil viewing camera, also driven by a Newport Controller. The purpose of this camera is to calibrate the position of the cold stop. Once this has been established, the camera is withdrawn from the light path and not used for subsequent operations.

The Pupilview Assembly will have two functional groups; one for deploying the pickoff optic, and the other for controlling the camera. The camera itself is discussed further in Section~\ref{sec:hxrg}.

% IFS SCALE

\subsubsection{IFS Scale (ifs.scale) and Resolution (ifs.res) Selectors}
\label{sec:ifs}

The IFS Scale Assembly (ifs.scale) configures the spatial resolution per spectral element (spaxel) of the spectrograph. The IFS offers two spectrograph channels: one using a lenslet array, and the other a slicer. The lenslet channel covers the finer plate scales of 4 and 9\,mas, while the slicer channel accommodates the coarser scales of 25 and 50\,mas. The main responsibility of the IFS Scale Assembly is to select one of these four scales. 

%\begin{figure}[ht]
%  \begin{center}
%  \includegraphics[width=0.7\linewidth]{pick_off_mirror_function.png}
%  \end{center}
%  \caption{\label{fig:ifs_pickoff}
%  IFS pickoffs.
%  }
%\end{figure}

There is a fixed fold mirror upstream of the imager detector that sends a portion of the light at the center of the imager field-of-view into the Lenslet IFU. Switching between Lenslet IFU and Slicer IFU is accomplished using the pick-off mirror that intercepts light in front of that mirror (near the bottom-left of Figure~\ref{fig:blockDiagram}), diverting light to the Slicer IFU. When using the Slicer IFU, a periscope mirror (located near the bottom-right in that figure) also needs to be inserted to pick-off the light from the slicer collimator and feed it to the common element grating. The assembly ensures that the periscope mirror works in tandem with the slicer pick-off mirror. 

%Figure~\ref{fig:ifs_pickoff} illustrates the functionality of the pick-off mirrors.

%\begin{figure}[ht]
%  \begin{center}
%  \includegraphics[width=0.5\linewidth]{reimaging_optic.png}
%  \end{center}
%  \caption{\label{fig:ifs_reimaging}
%  IFS reimaging optics.
%  }
%\end{figure}

Next, two sets of re-imaging optics are required to feed the Lenslet IFU and the Slicer IFU. The lenses within the mechanisms can be switched between two positions, depending on the selected resolution (these lenses can be seen near the bottom-left and top-right of of the IFS in Figure~\ref{fig:blockDiagram}).

The design for ifs.scale divides mechanism control into three functional groups: one for the pickoff, another for the re-imaging optics, and a final functional group for the periscope, though it will be possible to configure all of the mechanisms simultaneously with a single top-level command to obtain a particular resolution.

% IFS RESOLUTION

%\subsubsection{IFS Resolution Selector (ifs.resolution)}
%\label{sec:ifsRes}

The IFS Spectral Resolution (ifs.res) assembly configures the spectral resolution of the IFS which includes selecting a dispersing optic on a Grating Turret that is common to both arms of the spectrograph. The assembly must also position either a lenslet or mirror slicer mask stage (depending on the requested scale, and hence spectrograph arm) upstream of the Grating Turret to mask a portion of the light as a function of the desired spectral resolution (i.e. broad band versus narrow band). All of these mechanisms are located near the centre of the IFS in Figure~\ref{fig:blockDiagram}.

The Grating Turret has two moving stages: one is the main stepper motor that rotates the turret to select the grating, with switches to verify which grating is in position.  The other is a plunger mechanism that pushes and holds the active grating in place. The turret wheel has 16 discrete positions while the plunger has two positions. Since these axes will not be moving continuously we are currently planning to use Galil controllers to drive them. It is crucial for the safety of the mechanism not to drive the wheel mechanism while plunger is engaged; the ifs.res assembly will handle this coordination.

Like the Scale Assembly, ifs.res will consist of three functional groups to allow independent motion (primarily to assist with assembly and testing), though commands that could cause damage to the Grating Turret (i.e., independent plunger and grating motion) will be rejected. Generally speaking a single high-level command will be used to safely move all three components into the desired locations to achieve a particular resolution.

% -----------------------------------------------------------------------------
% INFRARED DETECTORS
% -----------------------------------------------------------------------------

\section{Infrared Detectors}

IRIS will have four different types of infrared detectors. Unlike CCDs, infrared detectors can be read multiple times throughout an integration (often referred to as a ``ramp'', since the voltage that is read grows over time as a function of incident flux). This feature, combined with the fact that each ``read'' (or ``readout'') is an independent sample that includes this fixed noise component, makes it possible to increase the signal-to-noise ratio (SNR) by increasing the number of reads within a ramp. At the end of each ramp (a length of time chosen such that the detectors do not saturate), the detectors are reset, with a new realization of the fundamental noise. One exposure may include one or more ramps to achieve longer effective integration time (i.e., an ``exposure'' may include multiple ramps). This terminology is used throughout the discussion of all the detectors in this section.

The Science Cryostat will house three different varieties of Teledyne Hawaii HxRG detectors: four $4096\times4096$ pixel H4RG-10 sensors for the imager, and one $4096\times4096$ pixel H4RG-15 for the IFS (the ``-10'' and ``-15'' suffixes refer to the pixel sizes in $\mu$m). At the time of writing it is expected that the Pupil Viewing Camera will use a smaller $2048\times2048$ pixel H2, which has a very similar interface (both electronically, and software) to the H4RGs. We are planning to use multiple Astronomical Research Cameras (ARC)\footnote{http://www.astro-cam.com} Generation IV controllers (ARC controllers henceforth), though most of the design work to date is based on team experience with the earlier Generation III controllers\cite{smith2012}. A single, configurable detector HCD is under development to support these three detectors.

In contrast, each of the three arms of the OIWFS will be connected to $320\times256$ pixel C-RED One (CRED1) camera. This choice was made during the Final Design Phase as these cameras offer substantially higher readout rates (full-frame) than what is possible using the Preliminary Design Phase choice of a $1024\times1024$ pixel H1RG. Despite the smaller sensor of the CRED1, the ability to read the full sensor at frame rates up to the maximum required for NFIRAOS (800\,Hz) will both simplify the design (such high rates with the H1RG would only be possible within very small readout windows), and potentially \emph{increase} sky coverage (at least for shorter exposure times). These commercially-available devices use the Cameralink standard to connect to COTS frame grabbers. A detector HCD is being designed for the CRED1 based around the selected frame grabber.

Each of the HxRG and CRED1 readout and software systems could be developed into full papers in their own right. Furthermore, a number of details will be worked out during prototyping early in the Build Phase. We therefore provide only an overview of the technology choices anf software architecture in this section.

% 
% HAWAII DETECTORS
%

\subsection{Teledyne HAWAII-xRG Detectors (imager.detector, imager.odgw, ifs.detector, pupilview)}
\label{sec:hxrg}

\begin{figure}[ht]
  \begin{center}
  \includegraphics[width=\linewidth]{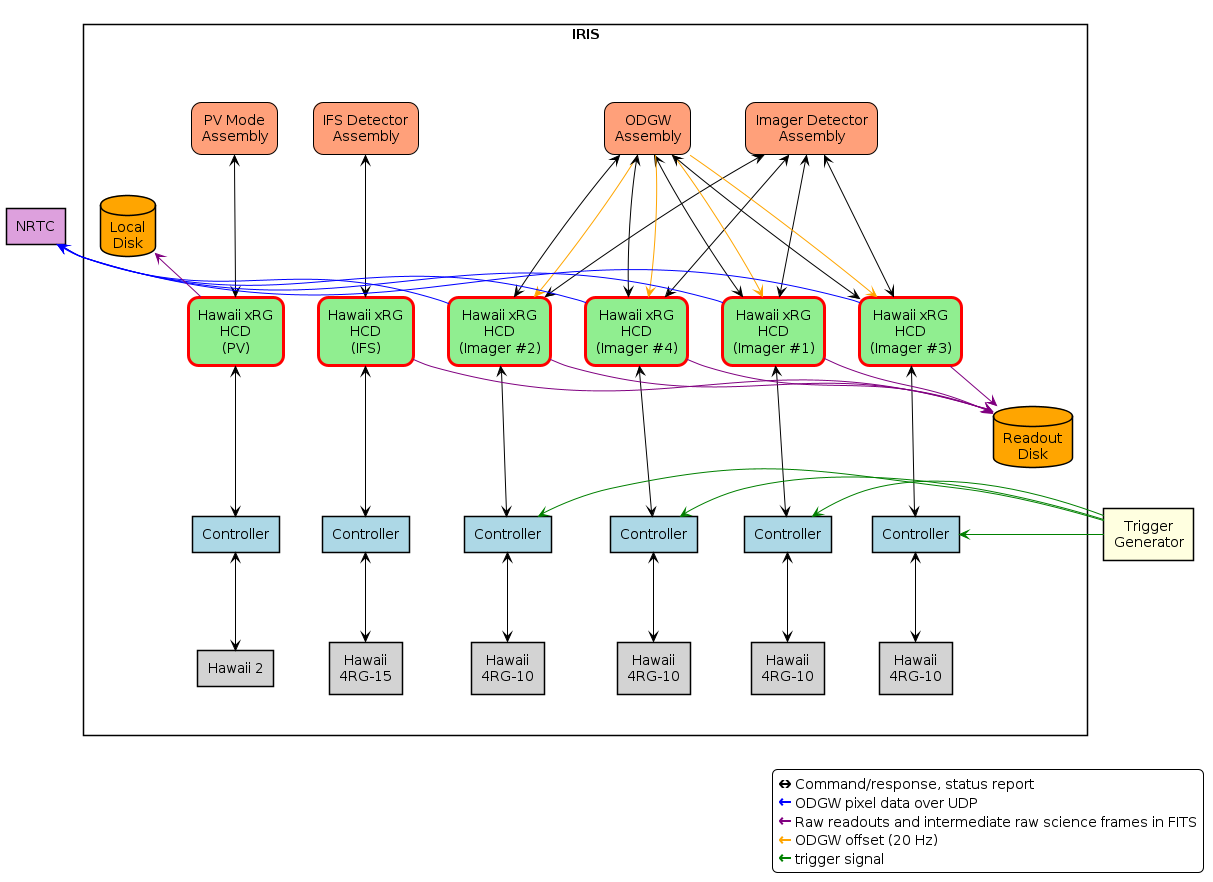}
  \end{center}
  \caption{\label{fig:hxrg_flow}
  HxRG HCD information flow. Note that the four Hawaii xRG HCD instances for the imager will be commanded both by the Imager Detector Assembly (for science exposures) and the ODGW Assembly (for on-detector guide windows). In all cases except the Pupil Viewing camera, detector reads will be written to a Readout Disk that will be accessible to the Data Reduction System (DRS).
  }
\end{figure}

The HxRG software stack must interface with TMT CSW services (using JVM-based clients), while low-level communication with the detector controllers is facilitated by a C/C++ API provided by ARC. Many of the software requirements are driven by the Imager Detector which needs to support the readout of ODGWs and ``subarrays'', interleaved with science exposures. ODGWs consist of up to four fast-readouts (at most one per imager chip) over small windows (e.g., several pixels on a side) that are used to measure tip/tilt information in the science focal plane as feedback for NFIRAOS. Like the OIWFS, the timing of these exposures must be precisely coordinated with the rest of the AO system, using an external external trigger from NFIRAOS to indicate when the exposures should occur, at rates of up to 400\,Hz. The ODGW pixels are transmitted to the NFIRAOS Real Time Controller (NRTC) from the detector host computer using a direct low-latency/high-bandwidth Ethernet connection. The HCD will run on a server with a real-time patched kernel to minimize timing jitter in the exposure, processing, and transmission of ODGW images. Like ODGWs, subarrays are small windowed readouts, though they are primarily intended to mitigate saturation of pixels in the vicinity of bright sources during long science exposures. Unlike ODGWs, there can be multiple subarrays per detector chip, and instead of sending the window data to the NRTC, they are only stored locally for use in the IRIS Data Reduction System (DRS) via a shared Readout Disk (probably an NFS mount). For both ODGWs and subarrays, the smaller window reads and resets are interleaved with full-frame reads, and the cadence of resets is of course much faster than for the full chip. The HCD will only allow either the ODGW or readout mode to be active at a time. Figure~\ref{fig:hxrg_flow} contrasts the information flow through the various instances of the HxRG HCD.

\begin{figure}[ht]
  \begin{center}
  \includegraphics[width=0.3\linewidth]{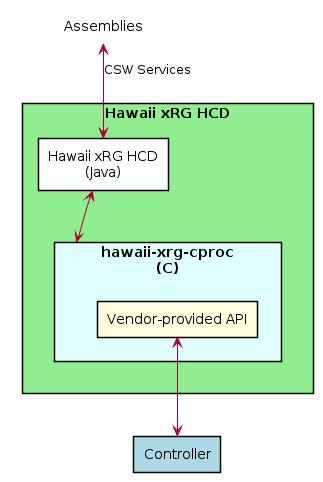}
  \end{center}
  \caption{\label{fig:hxrg_concept}
  Conceptual HxRG HCD software block diagram.
  }
\end{figure}

As shown in Figure~\ref{fig:hxrg_concept}, the HCD is actually envisaged as two separate applications: a Java program providing the TMT service interfaces, while the C/C++ layer is a separate process called ``hawaii-xrg-cproc''. The reason for this split, as opposed to embedding C function calls within the Java application using JNI or a similar approach, is to simplify the isolation and setting of real-time priority for hawaii-xrg-cproc, as well as the use of non-uniform memory allocation (i.e., ensuring that memory is attached to the CPU on which the application is running), without having to worry about JVM settings. This loose coupling also allows us to isolate CSW or Java-specific problems from the intrinsic functionality of the detector control. To minimize the volume of data transferred between those two layers, almost all of the core functionality will be implemented in hawaii-xrg-cproc, rendering the Java HCD a thin layer which only translates CSW commands and events into a form that hawaii-xrg-cproc can understand, and similarly translating responses and status updates from the C/C++ layer for CSW service clients. Note that if TMT provides a sufficiently full-featured set of C/C++ client interfaces to CSW (e.g., early in the IRIS build phase), we will consider combining both layers into one.

%\begin{figure}[ht]
%  \begin{center}
%  \includegraphics[width=0.9\linewidth]{pixel_data_flow-1.png}
%  \end{center}
%  \caption{\label{fig:hxrg_subarray}
%  HxRG HCD subarray mode: faster readouts of multiple subarrays are interlaced with lines of full-frame readouts to minimize the impact of bright sources.
%  }
%\end{figure}

For inter-process communication between the HxRG HCD (Java) and hawaii-xrg-proc (C) we are currently planning to use ZeroMQ to establish communication channels, and Google Protocol Buffers (GPB) for serialization. The advantage of this approach is that ZeroMQ sockets typically perform their I/O queuing in a background thread. This means that messages arrive in local input queues, and are sent from local output queues, no matter what our application is busy doing. This is desirable for the implementation of hawaii-xrg-cproc because the developer does not have to take care of the I/O blocking for communication, while processing realtime tasks. It is also good in the sense that communication is ``message'' based, unlike TCP streams or Unix Domain sockets. There is a clear boundary between each message, and the API provides a straightforward means for locating messages in the data stream. We note that these technology choices have also been made for other TMT software components, which should help to ease the build and long-term maintenance of the detector code described here.

For the configuration and control of the ODGW, the HCD will be commanded both by the ODGW Assembly (which is under control of the AO Sequencer), while science exposure and subarray configuration will be performed by the Imager Detector Assembly (which is commanded by the IRIS Sequencer). The HCD coordinates commands from both assemblies. For example, if the ODGW mode is active, a request to enable the subarray mode from the Imager Assembly will be rejected. The information flow for the four Imager detectors is compared with the simpler flow for the Pupil Viewing Camera and IFS detector in Figure~\ref{fig:hxrg_flow}, with additional details provided below.

%Figure~\ref{fig:hxrg_subarray} illustrates the flow of data and processing steps taken by the HCD (specifically the hawaii-xrg-cproc portion) for the full-frame readout %(top half of the diagram), and for subarrays (bottom half). Every raw readout (from either the subarray or the full frame line) is read from the conroller, de-scrambled, and saved as a FITS file. For the Imager and IFS detectors this storage is the Readout Disk, whereas it is a local disk in the case of the PV detector.

\begin{figure}[ht]
  \begin{center}
  \includegraphics[width=0.8\linewidth]{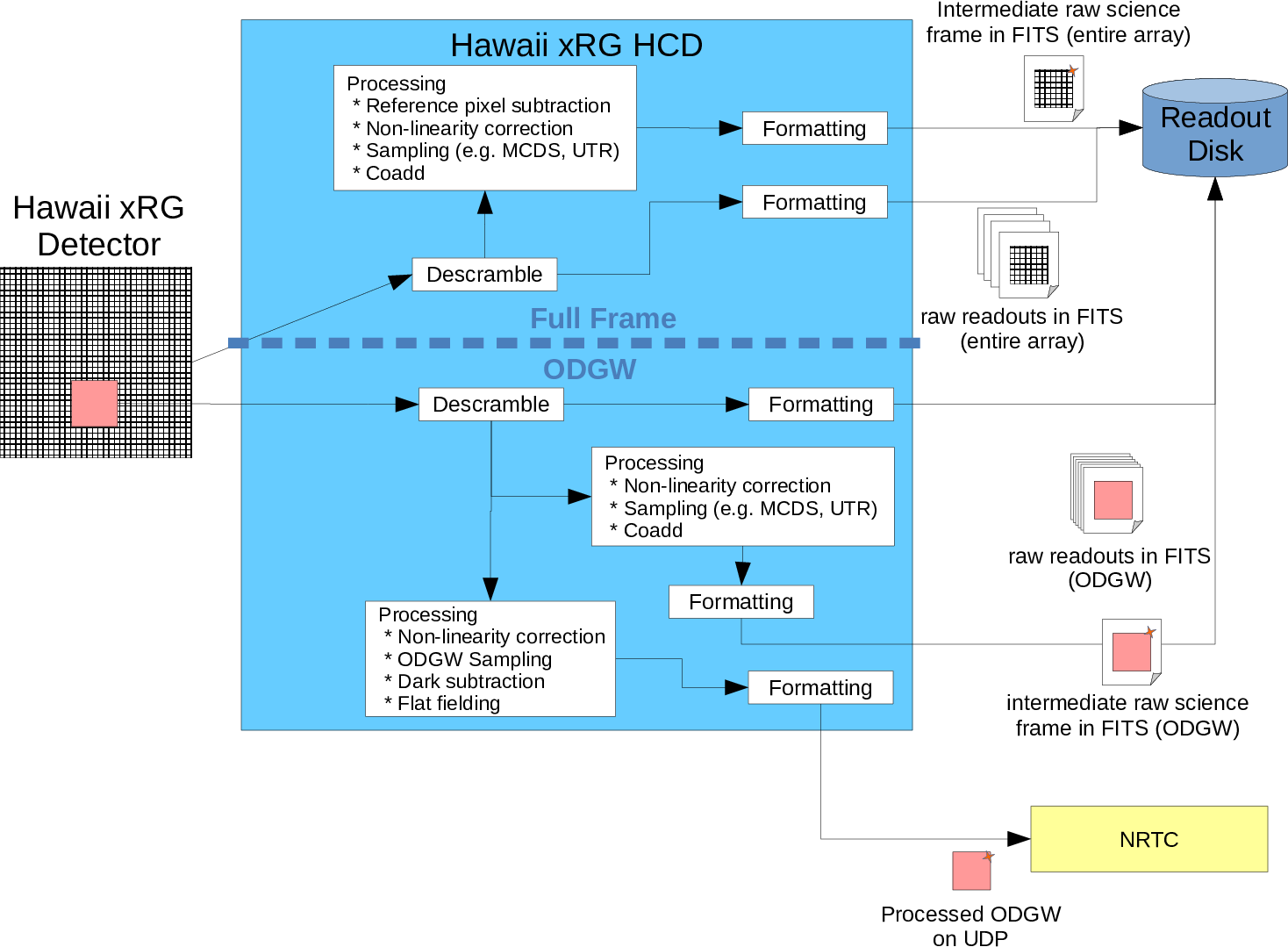}
  \end{center}
  \caption{\label{fig:hxrg_odgw}
  HxRG HCD ODGW mode: interlace full-frame readouts with faster windowed readouts (one per chip) to provide T/T feedback to AO system.
  }
\end{figure}

Figure~\ref{fig:hxrg_odgw} illustrates how the HCD works (specifically the hawaii-xrg-cproc portion), including full-frame readouts (top half) in parallel with an ODGW (bottom half). Every raw readout (from either the ODGW or the full frame path) is read from the controller, de-scrambled, and saved as a FITS file. For the Imager and IFS detectors this storage is the Readout Disk, whereas it is a local disk in the case of the Pupil View detector. For full-frame processing, the HCD will use multiple subsequent raw readouts to generate an ``intermediate raw science frame''. The steps include reference pixel subtraction, non-linearity correction, sampling arithmetic, and finally coadds. The sampling method will be either Multiple Correlated Double Sampling (MCDS) or Up-the-Ramp (UTR). The ODGW pixels will also undergo a similar level of processing. The actual algorithms will be implemented in a C library developed by the IRIS DRS team, to ensure that both this fast ``on-the-fly'' processing is consistent with processing performed by the DRS\cite{surya2020}. The raw reads, and the processed ODGW and full frame images are written as FITS files to disk.

In parallel, as shown at the bottom of the figure, there is a separate real-time path for the ODGW pixels. One or more threads with real-time priority perform non-linearity correction, Correlated Double Sampling (CDS), dark subtraction, and flat-fielding. These tasks are completed within the required frame processing time and sent over a dedicated link to the NRTC.

While not shown, the HCD will perform interleaved subarray processing in a manner similar to that of the ODGWs, with several key differences. First, while there can only be one ODGW per detector, there can be multiple windows in subarray mode. Second, ODGW positions can change throughout an observation at 20\,Hz (i.e., to support non-sidereal tracking), while subarrays are static throughout an observation. Finally, the timing of ODGW exposures is set by the external trigger, whereas no trigger is required in the subarray (or simpler full-frame science) case unless any one of the four detectors has an ODGW. Finally, when running in subarray mode, the bottom branch of the HCD showing pixel processing and transmission to the NRTC is omitted.

All of the intermediate raw science frames shown in the figure (written to the Readout Disk) are further processed by the DRS, and sent to the observatory for quick-look and permanent storage. They will also be useful for testing purposes. However, separately, the DRS will work offline from the stored raw readouts to produce science quality frames\cite{surya2020}.

%
% OIWFS DETECTOR
%

\subsection{OIWFS detectors (oiwfs.detector)}
\label{sec:oiwfsdetector}

\begin{figure}[ht]
  \begin{center}
  \includegraphics[width=0.7\linewidth]{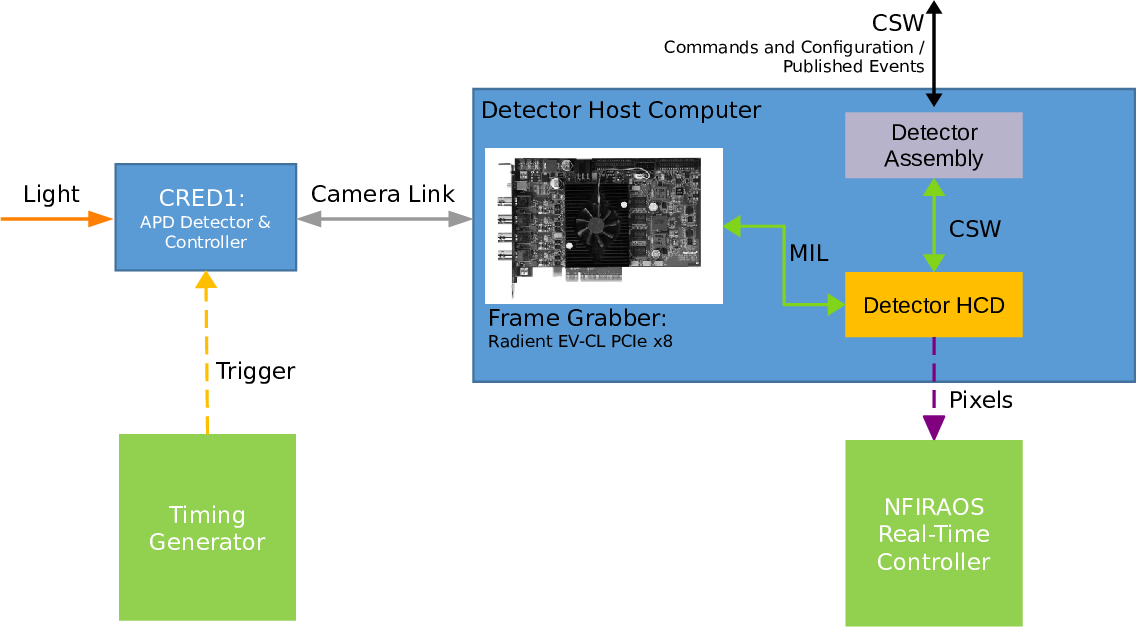}
  \end{center}
  \caption{\label{fig:cred1_context}
  Context diagram showing the CRED1 camera, the external trigger, computer, and communication interfaces.
  }
\end{figure}

The OIWFS will make use of three C-RED One (CRED1) infrared cameras produced by First Light Advanced Imagery\footnote{https://www.first-light-imaging.com/product/c-red-one/}, one for each probe. These cameras combine a SAPHIRA $320 \times 256$ pixel infrared Avalanche Photo Diode (APD) sensor, capable of taking up to 3500 images per second at full resolution, with an integrated controller that supports configuration and data transfer over Camera Link. Sensor data are digitized as 16-bit unsigned values. The camera will normally be placed into a continuous mode, with individual exposures triggered by an external signal from the AO system when guiding. Alternatively, the HCD will be able to self-trigger (e.g., for taking calibration exposures). A host computer uses a Camera Link-capable frame grabber card to interface with the camera. The HCD described here is intended to run on the host computer to provide a TMT CSW interface for the OIWFS Detector Assembly for the frame grabber, implementing all of the specific C-RED One configuration options and features required by the OIWFS. The OIWFS Detector Assembly may also run on the same server as the HCD (as in the figure below), but this is not a requirement.

The vendor currently supplies the cameras with a Matrox Radient eV-cl full frame grabber, which we plan to use in the final system. In addition to supporting the download of image data, the Camera Link connection includes an embedded serial line to configure the CRED1 (e.g., exposure mode parameters, uploading gain and bias frames for on-board image processing etc.). Matrox produces a Software Development Kit (SDK) of C-callable library functions called the Matrox Imaging Library (MIL), with native Linux support. The CRED1 HCD will make extensive use of the MIL to configure the CRED1 and to grab images.

The context diagram in Figure~\ref{fig:cred1_context} shows how the host computer and software communicate with the CRED1. The approach for the CRED1 HCD will closely match that of the HxRG HCD as discussed in Section~\ref{sec:hxrg}: there will be a lower-level standalone application written in C/C++ that can have the realtime priority configured, and it will interact with the frame grabber using the MIL. A thin Java layer will communicate with this application to expose a TMT CSW interface for the OIWFS Detector Assembly. We also plan to use a similar ZeroMQ and Google Protocol Buffers solution for inter-process communication. Unlike the HxRG HCD which will perform all data processing in software, the CRED1 supports on-board Correlated Double Sampling (CDS), bias subtraction, and flat-fielding. While the precise behavior will have to be studied when the cameras are delivered, it would certainly assist in minimizing latency in the OIWFS. Additionally, the images from the CRED1 will be cropped prior to their being sent to the NRTC to minimize network transfer latency. The locations of the cropped windows may be varied over time to support the on-chip guiding modes described in Section~\ref{sec:oiwfsPoa}.

% -----------------------------------------------------------------------------
% OTHER COMPONENTS
% -----------------------------------------------------------------------------

\section{Other Components}
\label{sec:other}

% OIWFS ENV
\subsection{OIWFS Environmental Control (oiwfs.env)}
\label{sec:oiwfsEnv}

Both the OIWFS and NFIRAOS are sealed, and cooled to -30 C using cooling panels for operations. While cooling or cold, outer skin heaters prevent exterior condensation. To avoid formation of frost, dry air is fed into NFIRAOS, and a slight positive pressure ensures that humid atmospheric air does not enter (when IRIS is not connected to NFIRAOS it can also operate with its own dry air purge). To improve convection while cooling and warming, the enclosure has air circulation fans. An optical bench heater is activated when warming to reduce warmup times. The interface between IRIS and NFIRAOS has a rotating labyrinth seal. ``Gate valves'' at both ends of the ``snout'' (an upward protrusion of the OIWFS into NFIRAOS), one controlled by NFIRAOS, and the other by IRIS, are used to protect their respective interiors, particularly if either (or both) instruments are cooling down or warming up. While cooling, heaters in the snout and on the faces of the gate valves ensure that any moisture is evaporated, while dry air purges remove this moisture. When both instruments have finished cooling to -30\,C, their two gate valves are opened, allowing the air to mix, and for light to pass through into IRIS. While operating (but not during observations), heating panels will be activated periodically within the OIWFS enclosure to remove any frost that may have formed, on a schedule that is dictated by NFIRAOS. The OIWFS enclosure also includes various refrigerant loops with remotely controlled valves to manage flow.

These components will be controlled and monitored using an Allen Bradley ControlLogix 1756\footnote{https://www.rockwellautomation.com/en-us/products/hardware/allen-bradley/i-o/chassis-based-i-o/1756-controllogix-i-o.html} or similar Programmable Logic Controller (PLC). These are the controllers currently recommended by TMT, and this model is of a similar type to the one that will be used for the NFIRAOS Optical Enclosure. Since the OIWFS PLC will resemble that of NFIRAOS, and since it will likely be implemented by the same group at NRC, it is expected that the software interface (HCD) will be similar (possibly sharing code). The main difference between the OIWFS enclosure and the NFIRAOS Optical Enclosure is that the latter is much more complex. Whereas NFIRAOS includes an air handling unit (AHU), must consider the safety of people entering the enclosure to perform maintenance, and also the states of multiple client instruments, etc., the OIWFS enclosure has a much simpler set of tasks to consider. To assist the controller in making decisions, it will have a number of thermometers, pressure, and humidity sensors mounted throughout the enclosure.

%\begin{figure}[ht]
%  \begin{center}
%  \includegraphics[width=0.5\linewidth]{oiwfsEnvStates.pdf}
%  \end{center}
%  \caption{\label{fig:oiwfsEnvStates}
%  OIWFS Environmental Control State Machine
%  }
%\end{figure}

The basic state machine model that we will implement in the PLC is summarized by the following points:

\begin{itemize}

  \item \textbf{WARM:} The enclosure is warm (ambient). All of the cooling loops (panels, optical bench), air  circulation fans, outer skin heaters, and dry air purges active, and the snout gate valve is closed.
  
  \item \textbf{WRMG:} The enclosure temperature is below ambient and warming. The bench heater, air circulation fans, and dry air purges are active, and the snout gate valve is closed.
  
  \item \textbf{COLD:} The enclosure temperature is -30\,C. The cooling panels and outer skin heaters are running to maintain the temperature.
  
  \item \textbf{CLNG:} The enclosure temperature is above -30\,C and is cooling. All of the cooling loops are active, air circulation fans, outer skin heaters, and dry air purges active, and the snout gate valve is closed.

  \item \textbf{CDPD:} Cooling panel defrost while the enclosure is at -30\,C. Note that it is up to the assembly to decide whether NFIRAOS is in an acceptable state prior to requesting a defrost.

\end{itemize}

The oiwfs.env assembly will provide and interface to the HCD, allowing users to request state changes. It will also monitor state events published by the NFIRAOS enclosure to determine when it is safe to perform a panel defrost (if not, transitions to the CDPD state will not be allowed). Finally, the assembly will monitor a number of other environmental sensors (particularly thermometers) on a 1Wire bus (via an 1Wire HCD). These temperatures will be published by the assembly for the benefit of the motorized mechanisms in the OIWFS. If a component is too hot, the respective component assembly will raise an alarm, and potentially also prohibit further use of the component until it cools (or is explicitly overridden). Since the 1Wire monitoring is completely passive, only a single functional group is required for oiwfs.env assembly for the PLC HCD interactions.

% SCIENCE CRYOSTAT ENV

\subsection{Science Cryostat Environmental Control (sc.cryoenv)}
\label{sec:cryoenv}

The Cryostat Environment Assembly serves multiple purposes, and has several functional groups: Imager, IFS, Window, Pupil Viewer, Pressure. This modular structure will be helpful through the many different phases of the project, especially when two halves of the cryostat are being assembled in two different locations. Each of Functional Groups will have different parameters to monitor, control, and raise alarms based on which thermal-vacuum state the cryostat is in (all specified in configuration files).  The top half of the cryostat, housing the Imager detectors, optics, and mechanisms, will have an Imager functional group. The bottom half has the IFS detector, optics and mechanisms, controller by the IFS functional group. The Window group controls the cryostat window heater (driven to the OIWFS enclosure temperature) to prevent frost formation, and to prevent air turbulence from affecting image quality. The Pupil Viewer group manages the pupil viewing camera CCD temperature (whose requirements differ from that of the imager), while in operation, and when stowed. Finally, the Pressure group monitors an Edwards TIC to monitor pressure changes during pumping, pressurizing, and also smaller variations during operations. In general, closed-loop heater control will be accomplished using Lakeshore 336s (using mechanism-mounted thermometers for feedback), with additional temperature monitoring via a Lakeshore 218 (possibly using a shared HCD that supports both device). Upon the initialization of the assembly and successful connection with HCDs for all of these components, each functional group will immediately begin periodic monitoring and logging.

A major function of the assembly is to manage cryostat environmental state transitions. Changing between the ambient environment and cryogenic vacuum must be carefully controlled and monitored. The following is a list that summarizes those states. Note that with the exceptions of WARM and COLD, relevant assemblies will limit, lock or turn off mechanisms and detectors that could be damaged while in the remaining intermediate states.

\begin{itemize}
  \item \textbf{WARM:} The instrument is at ambient temperature and pressure. Some warm testing of mechanisms and detectors is permitted. It can arrive at this state automatically when the target pressure is reached following the PRESSURIZING state,
  \item \textbf{PUMPING:} Air is being pumped out of the cryostat. During this time the instrument remains at the ambient temperature. The state can be entered from WARM.
  \item \textbf{WARM\_VACUUM:} The target vacuum has been achieved, but the instrument is still at the ambient temperature. This state can be reached either from PUMPING or WARMING.
  \item \textbf{COOLING:} The instrument is actively cooling. This state can only be entered from WARM\_VACUUM.
  \item \textbf{COLD:} The instrument is at the operational temperature and pressure. The state is entered automatically upon completion of COOLING.
  \item \textbf{WARMING:} Warming is commenced by deactivating the refrigeration. The state is entered from COLD.
  \item \textbf{PRESSURIZING:} A pressure valve is opened to allow the cryostat to return to atmospheric pressure. This state can be entered from WARM\_VACUUM.
\end{itemize}

% ELECTRONICS RACK

\subsection{Electronics Rack (el.env, el.power)}
\label{sec:elEnv}

An electronics rack on the Nasmyth platform hosts many of the IRIS computers and components, and has two assemblies associated with it: el.env for monitoring rack the temperature using a combination of 1Wire and Lakeshore 218 sensors; and el.power for providing access to Eaton IPC remote power bar (requiring a new HCD) for activating and deactivating all of the IRIS components.

% SAFETY SYSTEM

\subsection{Safety System}
\label{sec:safety}

At TMT, no software can be the primary mitigation for hazards. That responsibility instead falls to the Safety System, which will use a PLC to trigger interlocks when necessary. We plan to write a read-only HCD that monitors the state of this PLC, and publishes it as an event for the benefit of the IRIS assemblies. They will then use this information to provide meaningful error messages in the event that a component becomes interlocked (e.g., ``Rotator is no longer active due to Safety Interlock'').

% -----------------------------------------------------------------------------
% SUMMARY
% -----------------------------------------------------------------------------

\section{Summary and future work}

Along with the rest of the instrument, the IRIS control software design is now mature, and is approaching a Final Design Review in early 2021. We have demonstrated how to decompose this software into an Instrument Sequencer, and a series of assemblies and hardware control daemons for individual components, as required for the TMT observatory software architecture. We have a plan to produce a core library upon which these components will be built, to standardize their structure, approach to simulation and configuration, and to provide common naming conventions for things like commands, states, and telemetry. The software for some of the more complex components such as infrared detectors will be prototyped once the required hardware is procured early in the build phase, to minimize risk to the project. 

\acknowledgments

The TMT Project gratefully acknowledges the support of the TMT collaborating institutions.  They are the California Institute of Technology, the University of California, the National Astronomical Observatory of Japan, the National Astronomical Observatories of China and their consortium partners, the Department of Science and Technology of India and their supported institutes, and the National Research Council of Canada.  This work was supported as well by the Gordon and Betty Moore Foundation, the Canada Foundation for Innovation, the Ontario Ministry of Research and Innovation, the Natural Sciences and Engineering Research Council of Canada, the British Columbia Knowledge Development Fund, the Association of Canadian Universities for Research in Astronomy (ACURA) , the Association of Universities for Research in Astronomy (AURA), the U.S. National Science Foundation, the National Institutes of Natural Sciences of Japan, and the Department of Atomic Energy of India.

% References
\bibliography{refs} % bibliography data in refs.bib
\bibliographystyle{spiebib} % makes bibtex use spiebib.bst

\end{document}